\documentclass[doublecol]{epl2}
\usepackage{graphics,bm}
\usepackage{graphicx}
\usepackage{amsmath, amssymb, graphics}

\newcommand{\beq}{\begin{equation}}
\newcommand{\eeq}{\end{equation}}
\newcommand{\bqa}{\begin{eqnarray}}
\newcommand{\eqa}{\end{eqnarray}}
\def\gsim{\mathrel {\vcenter {\baselineskip 0pt \kern 0pt
\hbox{$>$} \kern 0pt \hbox{$\sim$} }}}
\def\lsim{\mathrel {\vcenter {\baselineskip 0pt \kern 0pt
\hbox{$<$} \kern 0pt \hbox{$\sim$} }}}

\protect

\title{Directional flow of solitons with asymmetric potential wells: Soliton diode}
\author{M. Asad-uz-zaman and U. Al Khawaja}
\institute{ Physics Department, United Arab Emirates University,
P.O. Box 15551, Al-Ain, United Arab Emirates.}


\abstract{ We study the flow of bright solitons through two
asymmetric potential wells. The scattering of a soliton by certain
type of single potential wells, e.g., Gaussian or Rosen-Morse, is
distinguished by a critical velocity above which solitons can
transmit almost completely and below which solitons can reflect
nearly perfectly. For two such wells in series with certain
parameter combinations, we find that there is an appreciable
velocity range for which solitons can propagate in one direction
only. Our study shows that this directional propagation or diode
behavior is due to a combined effect of the sharp transition in the
transport coefficients at the critical velocity and a slight
reduction in the center-of-mass speed of the soliton while it
travels across a potential well.}

\begin{document}
\maketitle

\section{Introduction}
\label{introduction section}

The interest in the problem of solitons scattering by external
potentials such as barriers
\cite{bar1,bar2,bar3,bar4,bar5,bar6,bar7}, wells
\cite{vaz91,well1,well2,well3,well4,brand1,brand2}, steps
\cite{steps1,steps2,steps3}, and surfaces \cite{surf1,surf2}, stems
from the nonlinearity involved and the wave nature of solitons
characterizing such process. For instance, the nonlinearity in the
Schr$\rm \ddot o$dinger equation that controls the evolution of
solitons, leads to the nonclassical behavior of partial
transmittance and partial reflectance upon scattering a soliton by a
potential barrier. The wave nature of solitons results also in a
host of other nonclassical scattering outcomes. Of particular
importance is the so-called quantum reflection \cite{qr}. It has
been shown that for slow enough solitons, complete reflection from a
potential well \cite{brand1} or a down potential step takes place
\cite{steps3}. Quantum reflection is caused by a transport of matter
or energy from the incident soliton to the bound state of the well
which interacts repulsively (attractively) with the original soliton
if these are out of phase (in phase). Furthermore, it was found that
for a certain type of single potential well, e.g., Gaussian or
Rosen-Morse (RM), the two scattering regimes, namely complete
quantum reflectance and complete transmittance, are separated by a
sharp transition taking place at a critical velocity \cite{brand1},
and this interesting phenomenon has recently been explained for
other systems using a delta potential \cite{wei12,wan12}.

Another important effect is the reduction in the center-of-mass
speed of the soliton as it crosses the potential region. This is due
to the fact that transmittance is not complete; a small fraction of
the soliton is reflected while most of the soliton is transmitted.
The reflected part is typically dispersive and decays in the form of
radiation. The larger fraction, i.e., the transmitted part, retains
its soliton shape and propagates but with some collective
oscillations. The energy for this kind of internal excitation comes
from the center-of-mass kinetic energy. As a result, the
center-of-mass motion of the soliton slows down.

These two effects, the criticality and the velocity reduction, can
be combined in a useful way when two or more RM potentials are
placed in series and can lead to novel phenomena. In this letter, we
study the dynamics of a soliton in the presence of two RM potential
wells with slightly different parameters. One of the interesting
effects of this potential combination is the directional flow of
solitons. For certain parameter combinations, there is a velocity
window for which solitons can propagate freely with nearly full
transmission in one direction and blocked with almost perfect
reflection in the other direction. We termed this behavior as
``soliton diode''. Asymmetric electron flow with two tunneling
barriers was studied \cite{wil91, jon92, oun02, hil06} and for
solitons this behavior has recently been demonstrated for a discrete
nonlinear Schr$\rm \ddot o$dinger equation with asymmetric
potentials \cite{italian}. While we obtain the transport
coefficients for the soliton diode numerically by solving the
appropriate Gross-Pitaevskii (GP) equation, we support our
qualitative explanation of this effect by investigating in detail
the dependence of the critical velocity and the velocity reduction
for each potential well separately in terms of the parameters of the
wells and the speed of the incident soliton. This allowed for an
independent and accurate account of the velocity window for which
this diode functions.

\section{Theoretical Model}
\label{theory}

In the presence of an external potential $V(x)$ the dynamics of
bright soliton is governed by the one-dimensional GP equation, which
can be written in standard dimensionless (for explicit units, see,
for example, Ref.~\cite{gor06}) form as
\begin{equation}
i {\partial\over\partial
t}\psi(x,t)=\left[-{1\over2}{\partial^2\over\partial x^2}+ V(x)
+g|\psi(x,t)|^2\right]\,\psi(x,t) \label{gpe},
\end{equation}
where we choose the potential $V(x)$ of the form
\begin{equation}
V(x) = - {V_1}\,{{\rm sech}^2\left[\alpha_1 (x-x_{\rm cm1}
)\right]}- {V_2}\,{ {\rm sech}^2\left[\alpha_2 (x-x_{\rm
cm2})\right]}. \label{potend}
\end{equation}
Here, $V_{1,2}$, $\alpha_{1,2}$, and $x_{\rm cm1,2}$ determine the
depth, inverse width, and the position of the center of the first
and second potential wells, respectively. This potential, which is
known in the literature as the Rosen-Morse potential, belongs to the
class of reflectionless potentials which may transmit without
reflection in the linear regime. The factor $g$ denotes the
mean-field interaction strength and we take $g=-1$. In our numerical
simulation we prepare an initial state far away from the potential
region and propagate it in real time. We always choose our initial
wave function as the exact solution of the homogeneous version of
Eq.~(\ref{gpe}), namely
\begin{equation}
\psi(x,t=0)= {A}\,e^{i\,v(x-x_{\rm cm})}{{\rm sech} \left[A(x-x_{\rm
cm})\right]} \label{psi0},
\end{equation}
with $A=1$ and $x_{\rm cm}=x_0+v\,t$, where $v$ and $x_0$ are the
initial center-of-mass velocity and position, respectively. The
evolution time is taken long enough for the scattered soliton to be
far away from the potential. Then, we calculate the reflection,
trapping, and transmission coefficients. For the left-to-right
moving soliton with a single potential well, we define the
reflection, trapping, and the transmission coefficients as: $
R=(1/N)\, \int_{-\infty}^{-l} |\psi (x,t)|^2 \,dx$, $L=
(1/N)\,\int_{-l}^l |\psi (x,t)|^2 \,dx$, and $T=
(1/N)\,\int_{l}^{\infty} |\psi (x,t)|^2 \,dx$, where $l\approx
5/\alpha$ from the center of the well and $N$ is the normalization
of the soliton given by $ N=\int_{-\infty}^{\infty} |\psi (x,t)|^2
\,dx$. For right-to-left moving soliton, $R$ and $T$ are
interchanged but $L$ remains the same. For the present case of two
potential wells in series, we take $-l$ to be on the left of the
left well and $l$ to be on the right of the right well.

\section{Results: The Diode Behavior}
\label{result}

\begin{figure}[htb]
\includegraphics[width=8cm]{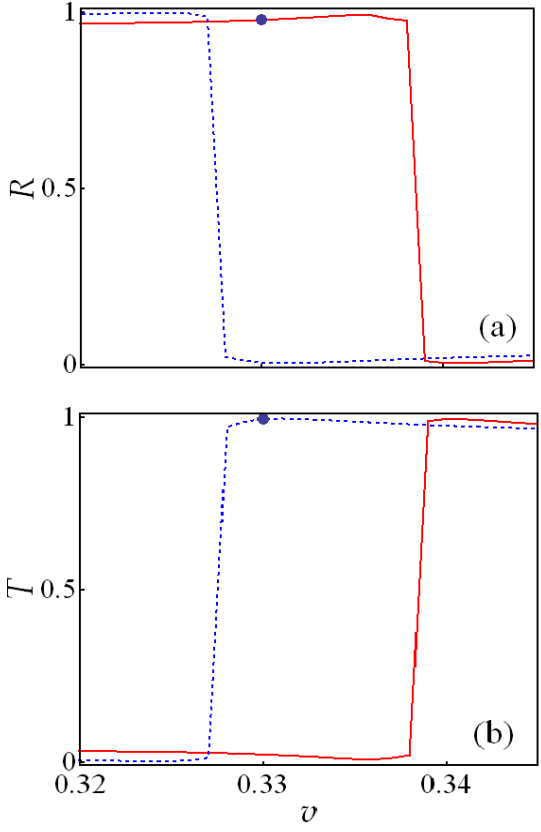}
\caption{(Color online) (a) Solid and dashed lines show the
reflection coefficients for the left and the right moving soliton,
respectively. (b) Solid and dashed lines show the transmission
coefficients for the left and the right moving soliton,
respectively. The parameter set is as follows: $V_1 = 4.35$,
$V_2=4.0$, $\alpha_1=\alpha_2=2$, and $x_{\rm cm1} = -x_{\rm
cm2}=6$. }\label{fig1}
\end{figure}

\begin{figure}[htb]
\includegraphics[width=8cm]{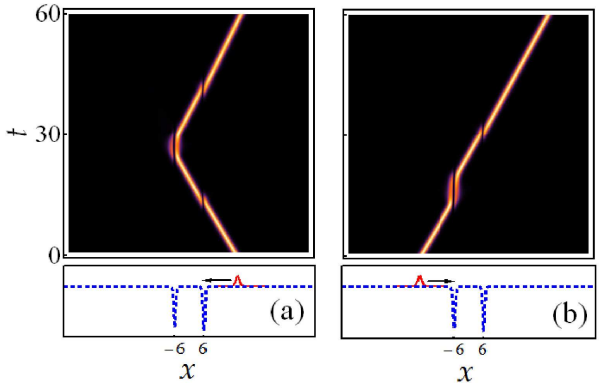}
\caption{(Color online) Figures (a) and (b) show the spatio-temporal
density plots with schematics for a case with $|v|=0.33$, which
corresponds to the dots in Figs.~\ref{fig1}(a) and (b),
respectively. The parameter set is same as in Fig.~\ref{fig1}
}\label{fig2}
\end{figure}

We present the main findings of our study here. We consider two RM
potential wells in series with slightly different parameters (See
the schematic figures in Figs.~\ref{fig2}(a) and (b) below.). For
this set of potential wells, the soliton is reflected for small
enough velocity and transmitted for large velocity. The transition
from high reflection to high transmission is very sharp.
Figures~\ref{fig1} (a) and (b) show the reflectance $R$ and
transmittance $T$ versus the magnitude of the incident soliton
velocity $v$ for the potential given by~(\ref{potend})  with $V_1 =
4.35$, $V_2=4.0$, $\alpha_1=\alpha_2=2$, and $x_{\rm cm1} = -x_{\rm
cm2}=6$. Solid and dashed lines in Fig.~\ref{fig1}(a) show the
reflection coefficient for the right-to-left moving and
left-to-right soliton, respectively. Figure~\ref{fig1}(b) shows the
transmittance for both cases. For the left-moving soliton there is a
critical velocity for which there is a sudden jump from mostly
reflectance to mostly transmittance. Similar behavior exists for the
right-moving soliton with another critical velocity. Surprisingly,
there is a difference in between the two critical velocities for the
left- and right-moving solitons. As a result, there is a small but
appreciable velocity window for which there is almost full
transmittance in one direction and nearly zero transmittance in the
other direction, i.e., the soliton shows directional propagation for
this set of parameters. This behavior is similar to the diode effect
in semiconductor physics. Figures~\ref{fig2}(a) and (b) show the
spatio-temporal plots corresponding to a selected value of the
initial speed. Figure~\ref{fig2}(a) corresponds to a soliton
incident from the right of the right well with incident velocity
$v=-0.33$. The soliton is first transmitted through the right well
(centered at $6$) and then reflected to the right by the left well
(centered at $-6$) and finally transmitted to the right through the
right well. The overall reflection coefficient is $R\approx 0.98$
and is marked by a filled circle in Fig.~\ref{fig1}(a).
Figure~\ref{fig2}(b) corresponds to a soliton incident from the left
of the left well and propagates to the right with the same magnitude
of the incident velocity $v=0.33$. This time the soliton is
transmitted through both wells. The overall transmission coefficient
is $T\approx 0.99$ and is also marked by a filled circle in
Fig.~\ref{fig1}(b).

This novel diode behavior of soliton propagation can simply be
explained by a combined effect of the critical velocity and the
reduction in the soliton's center-of-mass speed caused by a
potential well. Consider a soliton injected on a single potential
well with a speed slightly larger than the critical velocity. The
flow is naturally symmetric with full transmission from both sides.
Introducing another potential well with considerably lower critical
velocity, say to the right of the first one, breaks this symmetry.
While the soliton incident from the left transmits both the wells,
the soliton from the right with the same incident speed can only
transmit the well on the right since, as it crosses that well, it
suffers a velocity reduction which is small but enough to reduce the
soliton's speed to below the critical velocity of the left well.
Therefore, the soliton will be reflected from the left well to the
right. Since the critical velocity of the right well is still
considerably lower than the speed of the reflected soliton, the
soliton will continue its motion in the right direction. In this
manner, both the initially left-moving and the right-moving solitons
end up moving to the right.

As a specific example, we calculate, for the potential wells
considered in Figs.~\ref{fig2}(a) and (b), the critical velocities
of the left and right wells, $v_{cr}^l$ and $v_{cr}^r$,
respectively, which take the values: $v_{cr}^l=0.3275$ and $v_{cr}^r
= 0.1175$. Let us consider the case of the soliton with incident
velocity $v=0.33$. For this incident velocity, the velocity
reductions are $\Delta v^l =0.015$ and $\Delta v^r =0.014$. If it is
incident from the left, it will transmit through the left well as
the velocity is larger than $v_{cr}^l$. The transmitted soliton will
have the reduced velocity $v=0.315$  which is still larger than
$v_{cr}^r$, allowing the soliton to transmit through the right well
to the right. Now if the soliton is incident from the right side of
the right well, it will transmit through it, as the soliton's
velocity $v=0.33$ is larger than $v_{cr}^r$. Due to the velocity
reduction by the right well, the soliton's velocity reduces to
$v=0.316$ which is less than the $v_{cr}^l$, and therefore, the
soliton will be reflected by the left well to the right. Finally it
will be transmitted through the right well to the right. So whether
the soliton is incident from the right or from the left, it will
always end up moving to the right and there is no flow in the left
direction for the velocity considered.

Having demonstrated the function of the diode for a specific set of
the system's parameters and a single soliton's initial speed, a
thorough investigation of this phenomenon in terms of these
parameters is in order. Specifically, we want to account for the
velocity window in Fig.~\ref{fig1} over which the diode functions.
In addition, we want to find the effect of the potentials'
parameters on the width of this window with the aim of finding the
optimized potentials' parameters that lead to the maximum velocity
window.

\begin{figure}[htb]
\includegraphics[width=8cm]{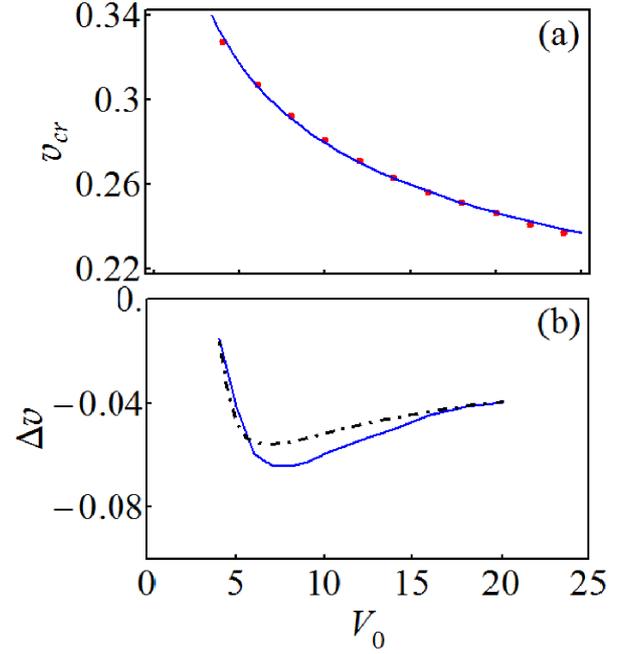}
\caption{(Color Online) (a) Critical velocity $v_{cr}$ as a function
of the well depth $V_0$ of the RM potential with
$\alpha=\sqrt{V_0}$. (b) Solid curve shows the velocity shift as a
function of the well depth $V_0$  of the RM potential with
$\alpha=\sqrt{V_0}$ for fixed incident velocity $v=0.33$.
Dash-dotted curve shows the velocity shift as a function of the well
depth derived semi analytically.} \label{fig3}
\end{figure}

We start by characterizing the effect of a single potential well,
$V(x) = -{V_0}\,{{\rm sech}^2( \alpha\,x)}$, on the propagation of
solitons. We found that it is essential to set $\alpha = \sqrt{V_0}$
in order to get a sharp transition of the transport coefficients at
the critical velocity. For such particular type of potential well,
the width of the well decreases with increasing the well depth. The
critical velocity turns out to be a monotonic decreasing function of
the well depth. Figure~\ref{fig3}(a) shows the critical velocity,
defined by $v_{cr}=( v_1+v_2)/2$ as a function of the well depth
$V_0$. Here $v_1$ is the largest velocity with almost full
reflection and $v_2$ is the smallest velocity with nearly perfect
transmission. Circles show our numerical data and the solid line
shows the fit of the data by $v_{cr}\approx0.42\,V_0^{-0.18}$.
Currently we are unable to find a convincing explanation to this
trend.

\begin{figure}[htb]
\includegraphics[width=6cm]{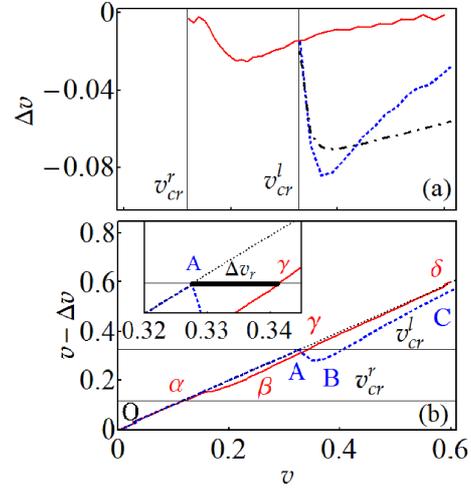}
\caption{(Color Online) (a) The velocity shift as a function of the
incident velocity. Solid curve shows the velocity shift for
$V_0=4.35$ and $\alpha = 2$ and dashed curve shows that for $V_0=4$
and $\alpha = 2$. Dash-dotted curve shows the velocity shift as a
function of the incident velocity derived semi analytically for
$V_0=4$. The vertical lines show the critical velocities for the two
well depths. (b) Reduced velocity versus incident velocity. The
horizontal lines show the critical velocities for the two well
depths. The solid and dashed curves show the reduced velocity for
the left and the right moving solitons, respectively. The inset
shows, with the thick horizontal line, the velocity range for which
the diode functions.} \label{fig4}
\end{figure}

Besides the critical velocity, it is also essential to calculate the
velocity reduction in terms of the system's parameters. The velocity
shift $\Delta v$ is defined as the difference between the incident
velocity and the velocity of the scattered soliton well after
crossing the potential region. This depends on the depth and width
of the potential well as well as on the magnitude of the incident
velocity. Figure~\ref{fig3}(b) shows the shift in velocity as a
function of the well depth $V_0$ for fixed incident velocity
$v=0.33$. This velocity is slightly larger than the critical
velocity $v_{cr}=0.3275$ for $V_0=4$ and $\alpha = 2$. The shift in
velocity also depends on the magnitude of the incident velocity. The
solid and dashed curves in Fig.~\ref{fig4}(a) show the shift in
velocity for incident velocities greater than the critical velocity
$v_{cr}$ and for fixed well depths $V_0=4.35$ and $V_0=4$,
respectively, with $\alpha = 2$. We calculate the velocity shift by
monitoring the position of the density maxima as function of the
evolution time. The slope of this curve gives the velocity and we
calculate two velocities one before the interaction with the
potential well and one after the interaction with the well. Both are
measured far from the interaction region.

To understand the physics of the diode's function, we employ a
variational calculation that confirms our previous arguments based
on the numerical results and specifically accounts for the velocity
reduction shown in Figs.~\ref{fig3} (b) and~\ref{fig4} (a).
Following Refs.~\cite{well1,brand1}, we use an ansatz function that
takes into account the incident soliton and a trapped mode, namely $
\Psi(x,t)= \psi_1(x,t)+\psi_2(x,t)$, where $\psi_1(x,t)= A_1\, {\rm
sech}[A_1\,(x-\eta_1)]\, {\rm exp}[i\phi_1+ i v_1 x]$ and
$\psi_2(x,t)= A_2\, {\rm sech}[x/q_1]\, {\rm exp}[i\phi_2+i\,
\sigma_2\, {\rm log\, cosh}(x/q_2)]$. Here $A_1$, $v_1$, $\eta_1$,
and $\phi_1$ denote the amplitude, velocity, center of mass
position, and the phase of the incoming soliton and $A_2$, $q_2$,
$\sigma_2$, and $\phi_2$ denote the amplitude, width, chirp
parameter, and the phase of the trapped mode, respectively. For
simplicity, the potential is approximated by the delta function
$V(x)=-V_0 \delta (x)$. After constructing the Lagrangian,
$L=\int_\infty^\infty dx\left[ \frac{i}{2}(\Psi^*
\frac{\partial}{\partial t}\Psi-\Psi \frac{\partial}{\partial
t}\Psi^*)- \frac{1}{2}|\frac{\partial}{\partial x}\Psi|^2
+\frac{1}{2}|\Psi|^4+ V|\Psi|^2\right]$, we derive the
Euler-Lagrange equations for the variational parameters. From the
set of resulting equations, we derive the following equation for the
center of mass position of the soliton
\begin{eqnarray}
\ddot{\eta_1}&=&
\dot{v_1}-\frac{d}{dt}\left(\frac{\dot{A_1}}{A_1}\right)\eta_1-
\left(\frac{\dot{A_1}}{A_1}\right)\dot{\eta_1},
\end{eqnarray}
where $\dot{A_1}/A_1=V_0\, A_2\, {\rm sech}(A_1\,\eta_1) \sin
(\Delta\phi)$ and $\dot{v_1}=-V_0\, A_1\, {\rm
sech}(A_1\,\eta_1)\tanh (A_1\,\eta_1)\left[A_2 \cos(\Delta\phi)
+A_1\,{\rm sech}(A_1\,\eta_1)\right]$ with
$\Delta\phi=\phi_1-\phi_2$ and $\dot{ }$ denotes the time
derivative. From this equation it is clear that the potential well
plays the role of a damping factor with a damping coefficient
$-{\dot A_1}/A_1$. Assuming $2\beta$ be the effective distance over
which the soliton experiences such a frictional force, the velocity
reduction can be shown to take the form
\begin{equation} \Delta v \approx -2 V_0 \beta A_1 A_2 {\rm
sech}(A_1\,\eta_1) \sin(\Delta\phi)\label{vr},
\end{equation}
which is proportional to the trapped mode amplitude $A_2$. If the
transmission is perfect, i.e., $A_1=1$ and $A_2=0$, there is no
velocity reduction. From this simple argument it is obvious that the
reduction in velocity arises primarily due to the imperfect
transmission after the critical velocity. In Fig.~\ref{fig1}(b), the
transmission coefficient curve acquires a slight downward curvature.
It is this particular behavior that mainly accounts for the velocity
reduction. Specifically, the coefficient of the trapped mode $A_2$
is proportional to $1-T$. To verify this, we have calculated $A_1$
and $A_2$ numerically in terms of the incident velocity and the well
depth. In addition, we approximated $\Delta\phi$ by its value near
the trap center, namely  $-.21\pi$ (See Fig.~7 in
Ref.~\cite{brand2}), and set $\eta=0$. Finally, we set $V_0
\beta=1/\sqrt{2 \pi}$, which is a good approximation for delta
potentials. As a result, we obtain the velocity reduction as a
function of the well depth and the incident velocity plotted with
dashed-dotted curves in Fig.~\ref{fig3}(b) and Fig.~\ref{fig4}(a),
respectively, which able to capture the main features qualitatively.

The mechanism by which the diode functions, can now be explained in
Figure~\ref{fig4}(b). The solid and dashed curves show the reduced
velocity of the soliton after crossing a single potential well for
the two cases of $V_0=4.35$ and $V_0=4$ with $\alpha = 2$. The
dotted diagonal line shows the velocity of the incident soliton. The
two horizontal lines mark the critical velocities of the two wells.
For the left-to-right-moving soliton we follow the dashed curve
marked as $OABC$. For $OA$, where the point $A$ corresponds to
$v=v_{cr}^l$, the incident velocity is smaller than the critical
velocity $v_{cr}^l= 0.3275$ of the left well and the soliton is
reflected back to the left. For $v> v_{cr}^l$ we follow the curve
$ABC$, where the reduced velocity is well above the critical
velocity $v_{cr}^r=0.1175$ of the right well and the soliton is
transmitted to the right. Therefore, from left to right, the soliton
is reflected for $v< v_{cr}^l$ and transmitted for $v
> v_{cr}^l$. On the other hand, for the right-to-left-moving soliton we
follow the solid curve marked as $O\alpha \beta \gamma \delta$. For
$O\alpha$, the incident velocity is smaller than the critical
velocity $v_{cr}^r$ of the right well and the soliton is reflected
back to the right. For $v
> v_{cr}^r$ we follow the line $\alpha \beta \gamma$,
where the point $\gamma$ corresponds to $v_{cr}^l+|\Delta v^r|$.
Here, the reduced velocity is well above the critical velocity
$v_{cr}^r$ of the right well but smaller than the critical velocity
$v_{cr}^l$ of the left well. As a result the soliton is reflected by
the left well and transmitted to the right through the right well.
For $\gamma\delta$ the reduced velocity is greater than the critical
velocity $v_{cr}^l$ of the left well and the soliton is transmitted
to the left. In conclusion, the right-moving soliton will transmit
to the right for $v>v_{cr}^l$ and the left-moving soliton will
reflect to the right for $v<v_{cr}^l+|\Delta v^r|$. As a result, the
scattered soliton will always be flowing to the right irrespective
of the direction of the incoming soliton provided its incident speed
is in the velocity range $v_{cr}^l<v<v_{cr}^l+|\Delta v^r|$. The
width of this velocity window $|\Delta v^r|\approx0.01$ agrees well
with that obtained directly from the numerical solution of the GP
equation, as can be seen by comparing Fig.~\ref{fig1}(a) and the
inset of Fig.~\ref{fig4}(b). It is essential to note that the
velocity reduction should be considerably larger than the velocity
range over which the transition at the critical velocity takes
place. If this condition is not met then partial transmission and
reflection, with comparable magnitudes, takes place resulting in a
failure of the function of the diode.

\begin{figure}[htb]
\includegraphics[width=8cm]{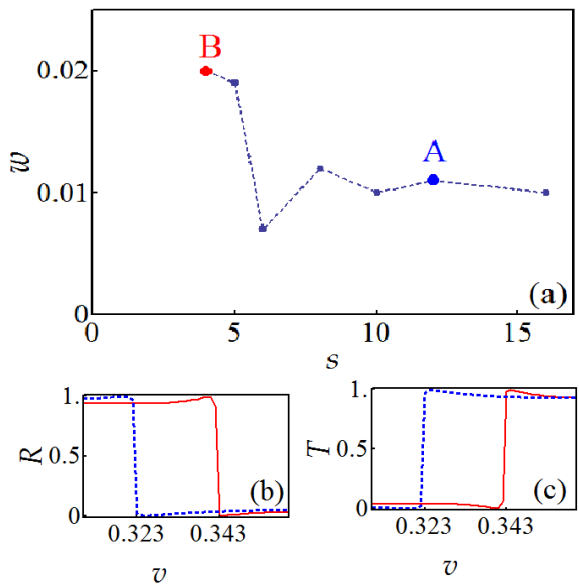}
\caption{(Color online) (a) Velocity window $w$ as a function of the
separation $s$ between the centers of the wells. The width and depth
of the wells are fixed. Dashed line is drawn to guide the eye. The
points marked as $A$ and $B$ are the separations which are used for
Fig.~\ref{fig1} and Figs.~\ref{fig4} (b) and (c), respectively. (b)
Solid and dashed lines show the reflection coefficients for the left
and the right moving solitons, respectively. (c) Solid and dashed
lines show the transmission coefficients for the left and the right
moving solitons, respectively. Position of the well centers are
$x_{\rm cm1}=-x_{\rm cm2}=2$. All the other parameters are same as
in Fig.~\ref{fig1}} \label{fig5}
\end{figure}

When the two wells are well separated, the effect of the either of
them is independent of the other. As we move the wells closer, they
couple together after a certain small separation and start to affect
the soliton propagation in a combined manner. Due to this coupling
between the wells and nonlinearity of the soliton dynamics, there is
a non-monotonous change in the velocity window for which the diode
effect is observed. Figure~\ref{fig5}(a) shows the width of the
velocity window $w$ as function of the separation $s$ between the
wells. We define the window width $w$ as the difference between the
critical velocities of the solitons transmitting from opposite
directions. The separation $s$ is the distance between the centers
of the two wells. Figures~\ref{fig5}(b) and (c) show the reflectance
and transmittance as a function of the incident velocity of the
soliton. Note that velocity window is now almost doubled compared to
that shown in Figs.~\ref{fig1}(a) and (b).

The directional flow of soliton with asymmetric potential wells can
also be achieved with two Gaussian wells with appropriate parameter
combination. Gaussian potential supports both criticality
\cite{brand1} and velocity reduction for soliton propagation. For
example, the potential $V(x)=-4.35 \exp[{-3 (x-6)^2}]-4.0 \exp[{-3
(x+6)^2}]$ shows diode behavior in the range $0.304\leq v \leq
0.310$. Therefore, the directional flow is not unique to a series of
RM potentials.

The combined effect of the velocity reduction and criticality can
also be used to trap solitons. If the reduction in velocity changes
the velocity such that it is smaller than any of the two critical
velocities then the soliton will be trapped in between the potential
wells and oscillates. One such parameters combination is $v=0.33$,
$V_1=V_2=4$, $\alpha_1 =\alpha_2=2$, and $x_{\rm cm1}=-x_{\rm
cm2}=6$. For this set of parameters, if the soliton is incident from
left or from right it can transmit through the first well and the
reduced velocity becomes $v=0.315$, which is smaller than
$v_{cr}^l=v_{cr}^r=0.3275$ and oscillates between the two wells.

For multiple soliton cases, we find similar behaviors as for single
soliton case only if the initial soliton separation is very large
compared to soliton width. In this regard, we use a chain of two
solitons, which are initially located at, say, $x_{\rm cm}=20$ and
$x_{\rm cm}=60$. Qualitatively we get the same diode behavior as
shown in Fig.~\ref{fig1}. For smaller initial separation, the
nonlinearity and the interaction near the potential regions destroy
the diode effect.

\section{Discussion and Conclusions}
\label{concsec}

Our numerical study of the dynamics of one-dimensional GP equation
with two asymmetric potential wells (Gaussian or Rosen-Morse) shows
that the soliton can propagate in one direction for a certain
velocity range and with certain parameter combinations. This
unidirectional propagation arises due to the criticality and the
velocity reduction by a single potential well. When two such
potentials with slightly different parameters are in series, the
combined effect causes the soliton to move in one direction only. A
detailed account of the critical velocities and velocity reductions
of both potential wells explains this effect and predicts the
velocity range over which the diode functions, in a good agreement
with the numerical result. We also employ a variational calculation
and estimate the velocity reduction as a function of the input
velocity and the depth of the potential well. The semi analytical
velocity reduction curves able to capture the main feature of the
corresponding exact numerical ones qualitatively.

The function of the soliton diode was demonstrated here for a single
pair of potential depths $V_1=4.35$ and $V_2=4.0$. The width of the
velocity window for which the diode works properly, certainly
depends on the values and ratio of these two parameters as well as
on the widths of the potential wells and their separation. Thus,
widening the velocity window and increasing the diode's efficiency
requires further systematic investigation of the effect of these
parameters as well investigating other types of potential wells.

The GP equation considered here is the governing equation of the
dynamics of both matter-wave solitons in Bose-Einstein
condensates~\cite{hul02,sal02} and optical solitons in optical
fibers. Therefore, it is expected that the diode behavior found here
to be realized in both cases. For the case of optical solitons, this
diode may be a favorable instrument in achieving all-optical
communication and computation in optical fibers.

\acknowledgments
The authors acknowledge the support provided by
United Arab Emirates University under the UAEU-NRF 21S038 grant and
the support provided by King Fahd University of Petroleum and
Minerals under group project numbers RG1107-1, RG1107-2, RG1214-1,
and RG1214-2.

\end{document}